# Ab-initio study of mechanical and electronic properties of MoAlB

M. A. Ali[1], M. A. Hadi[2*], M. M. Hossain[1], S. H. Naqib[2] and A. K. M. A. Islam[2,3]

[1]Department of Physics, Chittagong University of Engineering and Technology, Chittagong-4349, Bangladesh
[2]Department of Physics, University of Rajshahi, Rajshahi-6205, Bangladesh
[3]International Islamic University Chittagong, 154/A College Road, Chittagong-4203, Bangladesh

**ABSTRACT**

Using first-principles calculations, the structural, elastic and electronic properties of MoAlB have been investigated for the first time. The optimized lattice constants exhibit fair agreement with the experimental values. The computed elastic constants satisfy the mechanical stability conditions for the MoAlB. The Mo-based boride MoAlB is elastically anisotropic and can be classified as brittle material. This boride is expected to show reasonable thermal conductivity due to its high Debye temperature of 693 K. The metallic electrical conductivity of this compound is predicted from the electronic band structure calculations. The chemical bonding in MoAlB is basically covalent in nature which is supported by the calculated electronic density of states (DOS), Mulliken population, and charge density distribution. The estimated hardness value of 11.6 GPa for MoAlB suggests that it is softer compared to many other borides. The Fermi surface is formed due to low dispersive Mo 4d-like bands, which makes the compound a conductive one.

## 1. Introduction

Recent studies on the ternary borides are paying attention mostly on single-crystal growth and on analyzing their crystal structures. For instance, Ade *et al.* [1] in the recent times synthesized single crystals of some earlier reported $M_2AlB_2$ (M = Cr, Mn, Fe) and MAlB (M = Mo, W) materials, and deliberated their structural relation to other borides of transition metals and their resemblance to the MAX phase compounds. M-site solid solutions, $(Mo_x,Me_{1-x})AlB$, with Me = Cr, W and $(Fe_2,Me_{2-x})AlB_2$, with Me = Cr, Mn, were synthesized in near past [2, 3]. In this continuation, Kota *et al.* [4] synthesized dense, pre-eminently single-phase MoAlB, via a reactive hot pressing technique. Their HRSTEM (High-resolution scanning transmission electron microscopy) study affirmed the existence of two aluminum-layers in between a molybdenum-boron sub-lattice. Even borides, which are some of the hardest and heat-resistant materials, oxidize at high temperatures, breaking down the structural integrity of the compounds. However, Kota *et al.* predicted that the behavior of their boride is markedly better than that of others' MoAlB [2, 5, 6] and have a protective layer of aluminum, which makes it world's first corrosion-resistant boride. Inspired by MAX compounds, Kota *et al.* speculated they could make MoAlB with a structure similar to MAX phases, which would assist the material to resist oxidation at high temperatures. At high temperatures, Al atoms diffuse to the surface of MoAlB, where the Al atoms react with O atoms to form a shielding layer of alumina. As a result, the material forms its own protective coating. In addition, MoAlB retains its high conductivity at elevated temperature, making it potentially useful in a series of applications. Its high melting temperature, greater than 1400°C, suggests that it may lead to growth of ultra-high-melting-point borides capable of oxidation resistance. This innovation is quite momentous, since it is the first example in the history of a transition metal boride that is resistant of oxidation. Moreover, MoAlB is a good thermal conductor (35 $Wm^{-1}K^{-1}$ at room temperature). Its resistivity is low (0.36 - 0.49 μΩm) at room temperature and like a metal, decreases with the decrease of temperature. This compound maintains its stability up to at least 1400°C in inert atmospheres. In the temperature range 25 - 1300°C, its thermal expansion coefficient is found to be $9.5 \times 10^{-6}$ $K^{-1}$. Relatively low Vickers hardness values of 10.6 ± 0.3 GPa, in comparison to other transition metal borides, and ultimate compressive strengths up to 1940 ± 103 MPa should make the compound technologically very important.

---

*Corresponding author: hadipab@gmail.com

To the best of our knowledge, no ab-initio study on structural, elastic, and electronic properties exist for MoAlB in the literature. This encourages us to study this compound theoretically using density functional methodology for the first time.

**2. Computational approaches**

The first-principle computations are carried out by using the pseudo-potential plane-wave method derived from the density functional theory (DFT) [7, 8] put into operation by the CASTEP code [9]. The exchange-correlation interaction is treated with the generalized gradient approximation (GGA) within the scheme of Perdew-Burke-Ernzerhof (PBE) [10]. Vanderbilt-type non local ultra-soft pseudo-potentials [11] are used to treat the tightly-bound core electrons. The states B $2s^2$ $2p^1$, Al $3s^2$ $3p^1$ and Mo $4p^6$ $4d^5$ $5s^1$ are treated as valence states. The two important parameters that control the accuracy of the calculations are the kinetic energy cut-off, which fixes the number of plane waves in the expansion and the number of special *k*-points that determines the Brillouin zone (BZ) sampling. The number of *k* points and the energy cutoff are increased step by step until the calculated total energy converges within the required level of tolerances. The well converged total energy is achieved with 8×2×8 special k-points mesh [12] and energy cutoff of 400 eV. The convergence tolerance is set to energy change below $5.0 \times 10^{-6}$ eV per atom, force less than 0.01 eV/Å, stress less than 0.02 GPa, and change in displacement less than $5.0 \times 10^{-4}$ Å. For smooth Fermi surface and charge density map 22 × 8 × 22 *k*-point mesh is used.

## 3. Results and discussions

### *3.1. Structural properties*

MoAlB crystallizes in an orthorhombic unit cell with three different lattice constants *a*, *b*, and *c*. Three kinds of atoms Mo, Al, and B all occupy the 4c Wyckoff positions of space group *Cmcm*-$D_{2h}^{17}$. The crystal structure of MoAlB is suitably described in terms of the trigonal prismatic array of six Mo atoms enclosing each B atom. Two B and one Al atom are found to be located outside the rectangular faces of the trigonal prism. The prisms are packed in such a way that all prism axes are parallel to the *b* direction, while the B atoms form B-B zigzag chains in the *c* direction. The Al atoms form firmly wrinkled metal layers interleaved between the Mo double layers [13]. The calculated lattice parameters along with corresponding experimental data are listed in Table 1. The theoretical results are highly consistent with the experimental values with the deviations not more than 0.88%. Therefore, no confusion should arise on the reliability of the present first-principles DFT calculations.

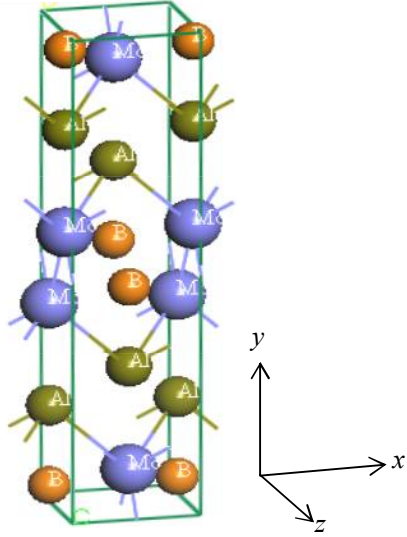

**Fig. 1.** Crystal structure of Mo-based ternary boride MoAlB.

**Table 1**. Structural parameters of ternary boride MoAlB.

| *a* (Å) | *b* (Å) | *c* (Å) | $y_{Mo}$ | $y_{Al}$ | $y_B$ | $V$ (Å³) | Ref. |
|---|---|---|---|---|---|---|---|
| 3.215 | 14.049 | 3.106 | 0.0890 | 0.302 | 0.467 | 140.28 | This |
| 3.213 | 13.986 | 3.103 | | | | 139.51 | [2] |
| 3.21 | 13.98 | 3.10 | | | | 139.11 | [4] |
| 3.209 | 13.980 | 3.100 | | | | 139.06 | [5] |
| 3.212 | 13.985 | 3.102 | 0.0891 | 0.304 | 0.465 | 139.34 | [6] |

## 3.2. Elastic properties

*Single crystal elastic constants*

The elastic constants link the response of the material to external forces, as characterized by bulk modulus, shear modulus, Young's modulus, and Poisson's ratio, and clearly take part in determining the mechanical strength of the materials. The elastic constants also give important information regarding the bonding characteristic between adjacent atomic planes and the anisotropic nature of the bonding and structural stability. For an orthorhombic crystal the most common 21 non-zero independent elastic constants reduce to nine components, i.e. $C_{11}$, $C_{22}$, $C_{33}$, $C_{44}$, $C_{55}$, $C_{66}$, $C_{12}$, $C_{13}$ and $C_{23}$ due to symmetry between stress and strain tensors. The elastic constants can be determined by computing the resulting stress generated due to applying a set of given homogeneous deformation of a finite value within the CASTEP code from first-principles method [14]. This method has been effectively used to predict the elastic properties of a series of materials including many metallic systems [15 – 24].

Necessary and sufficient Born criteria [25] for mechanical stability of an orthorhombic system are as follows:

$C_{11} > 0$; $C_{44} > 0$; $C_{55} > 0$; $C_{66} > 0$;

$C_{11}C_{22} > C_{12}^2$

$(C_{11}C_{22}C_{33} + 2C_{12}C_{13}C_{23}) > (C_{11}C_{23}^2 + C_{22}C_{13}^2 + C_{33}C_{12}^2)$

The calculated nine independent elastic constants for MoAlB are listed in Table 2, which satisfy the above mentioned stability criteria. Therefore, the ternary boride MoAlB is mechanically stable. The diagonal elastic constants $C_{11}$, $C_{22}$, and $C_{33}$ measure the resistance to linear compression along the crystallographic *a*, *b*, and *c* axes, respectively. The calculated value of $C_{22}$ is smaller than $C_{11}$ and $C_{33}$, which means that the ternary boride MoAlB is more compressible along the *b*-axis than that of along *a*- and *c*-axis. The off-diagonal, shear components of the elastic constants are $C_{12}$, $C_{13}$, and $C_{23}$. The constants $C_{12}$ and $C_{13}$ are almost equal. The elastic tensors $C_{12}$ and $C_{13}$ combine a functional stress component in the crystallographic *a* direction with a uniaxial strain along the crystallographic *b* and *c* axes, respectively. The large values of these elastic components would recommend that the crystal system MoAlB is capable to resist shear along the crystallographic *b* and *c* axes, when a large force is applied along the crystallographic *a* axis. Among three shear components $C_{23}$ has the lowest value. It combines a uniaxial strain along the crystallographic *c* direction to a functional stress component along the crystallographic *b* direction. The lowest value of $C_{23}$ reflects the observation of the large different between $C_{22}$ and $C_{33}$ in magnitude. This suggests a large discrepancy in the strength of the inter-molecular forces along either of these two directions. This non-equilibrium of forces in the different crystallographic axes could result a decreased resistance to shear, relative to the other off-diagonal tensor components. The elastic constant $C_{44}$ is the most significant parameter indirectly leading the indentation hardness of materials. The large value of $C_{44}$ indicates material's ability of resisting the shear deformation in (100) plane and the $C_{66}$ reflects the resistance to shear in the <110> direction. For comparison, the elastic constants and moduli of two silicide ternaries [26] are also given in Table 2. It is seen that all constants and moduli, excepting $C_{22}$ of TiPtSi, are large for MoAlB.

**Table 2.** Single crystal elastic constants $C_{ij}$ (in GPa), polycrystalline bulk modulus $B_V$, $B_R$, $B$ (in GPa), shear moduli $G_B$, $G_R$, $G$ (in GPa), Young's modulus $E$ (in GPa), Pugh's ratio $G/B$ and Poisson's ratio $v$ for MoAlB.

| Single crystal elastic properties | | | | Polycrystalline elastic properties | | | |
|---|---|---|---|---|---|---|---|
| Types | MoAlB [This] | ZrPtSi [26] | TiPtSi [26] | Types | MoAlB [This] | ZrPtSi [26] | TiPtSi [26] |
| $C_{11}$ | 349 | 305 | 286 | $B_V$ | 209 | 179 | 184 |
| $C_{22}$ | 320 | 308 | 341 | $B_R$ | 207 | 179 | 182 |
| $C_{33}$ | 399 | 341 | 353 | $B$ | 208 | 179 | 183 |
| $C_{44}$ | 190 | 143 | 135 | $G_V$ | 148 | 122 | 119 |
| $C_{55}$ | 160 | 125 | 125 | $G_R$ | 139 | 120 | 118 |
| $C_{66}$ | 169 | 133 | 122 | $G$ | 144 | 121 | 118 |
| $C_{12}$ | 141 | 113 | 114 | $E$ | 351 | 296 | 292 |
| $C_{13}$ | 146 | 104 | 104 | $B/G$ | 1.44 | 1.48 | 1.55 |
| $C_{23}$ | 118 | 112 | 120 | $v$ | 0.22 | 0.22 | 0.23 |

*Elastic constant and moduli are shown in round figures.

*Cauchy relations and Cauchy pressure*

To judge the anisotropy of the lattice potential the Cauchy relations [27] are used frequently. The elastic constants for the crystal systems that are centrosymmetric and firmly follow a central potential, should be completely symmetric in their four suffices. These symmetry relations for an orthorhombic space group are $C_{23} = C_{44}$, $C_{13} = C_{55}$, and $C_{12} = C_{66}$. For the MoAlB, $C_{23}$ is 37.89% smaller than $C_{44}$, $C_{13}$ is 8.75% less than $C_{55}$ and finally $C_{12}$ differs from $C_{66}$ by 16.57%. It may be said that all of these constants differ by significant amount, pointing that the Cauchy relations are far from applicable for MoAlB. This clear indication implies that many-body forces are essential to precisely predict the properties of MoAlB in the solid state.

The difference ($C_{12} - C_{44}$) is well known as Cauchy pressure. Pettifor [28] suggested that the Cauchy pressure can be used to describe the character of chemical bonding in solid materials. The positive value of ($C_{12} - C_{44}$) indicates the metallic bonding, whereas the negative value signifies the directional covalent bonding with angular character. The large negative value represents more directional characteristic of bonding. Therefore, the ternary boride MoAlB should exhibit the significant directionality in its covalent bond due to having a large negative value of –49. A positive Cauchy pressure always indicates ductile nature of a material, while a negative value corresponds to brittleness. Hence the ternary boride MoAlB should behave in a brittle manner.

*Bulk elastic properties*

Bulk and shear moduli of polycrystalline aggregates are calculated from individual elastic constants, $C_{ij}$, by the well-established Voigt [29] and the Reuss [30] approximations that are frequently used in averaging the single-crystal elastic constants for polycrystalline behavior in accordance with the Hill approximation [31]. According to Hill approximation, the bulk modulus $B$ and shear modulus $G$ are given by:

$$B = \frac{1}{2}(B_V + B_R) \text{ and } G = \frac{1}{2}(G_V + G_R)$$

where $B_V$, $B_R$ and $G_V$, $G_R$ are the Voigt and Reuss values for bulk and shear modulus, respectively. All calculated values are listed in Table 2. It is seen that the difference between $B_V$ and $B_R$ as well as $G_V$ and $G_R$ is comparatively small. According to Hill, the difference between these limiting values may be proportional to the degree of elastic anisotropy of crystal. Hence, the ternary boride should exhibit slight anisotropy in elastic behavior. Bulk modulus is of immense interest owing to its importance in evaluating the mechanical properties of solids. It reflects the ability of solids to resist compression. At the microscopic level, the bulk modulus depends on the nature of chemical bonds in solids. The comparatively high bulk modulus of MoAlB indicates its ability to defend against volume deformation and having strong chemical bonding. The shear modulus is a measurement of resistance to plastic deformation i.e., shape change in solids. The higher the shear modulus, the better the ability for a material to resist shearing forces and so, the more rigid the material is. It is more pertinent to materials' hardness and its large value is mainly due to materials' large $C_{44}$. Because of large value of shear modulus as well as $C_{44}$ the ternary compound MoAlB should have the ability to resists the shape change and a high level of hardness. Bulk and shear modulus are strongly related to the failure mode of solids. The failure of solids is typically categorized into brittle failure i.e., fracture and ductile failure i.e., plastic deformation. A material is either brittle or ductile for most practical situations. The bulk modulus to shear modulus ratio, $B/G$, well known as Pugh's ratio [32] is frequently used as a measure of brittle or ductile behavior of solids. The larger the $B/G$ value, the higher the ductility of solids and a critical value of $B/G$ is 1.75, which classifies a material to be ductile or brittle in nature. Specifically, if $B/G > 1.75$, the material should be characterized as ductile manner, otherwise the material exhibits brittle nature. From Table 2, it is noted that the ternary boride MoAlB should behave in a brittle manner.

Further, the Young's modulus $E$ and Poisson's ratio $v$ are determined from the following equations:

$$E = \frac{9BG}{3B + G}$$

and

$$v = \frac{3B - 2G}{6B + 2G}$$

The Young's modulus is a measure of the stiffness of solid materials. It also estimates the resistance against longitudinal stress. The larger the Young's modulus, the stiffer the material is. The Young's modulus of MoAlB is 351 GPa, which should put MoAlB under the category of a stiff material. The factor, which can predict the stability of crystal systems against shear, is the Poisson's ratio $v$. The relatively small Poisson's ratio is related to the systems stability against shear [33]. The estimated Poisson's ratio for MoAlB is 0.22. It is expected that this ternary boride should be stable against shear due to its low Poisson's ratio. The Poisson's ratio also predicts the character of inter-atomic forces in solid materials [34]. A material is said to be central force solid when its Poisson's ratio has values within 0.25 and 0.5, otherwise it is non-central force solid. It is evident that the stability is achieved from non-central forces in MoAlB, which is already predicted from Cauchy pressure. Further, Poisson's ratio is used often in engineering science, and it relates directly to the failure mode of solids. The critical value of $v$ is 0.26, which identifies a material either as brittle or ductile one [35, 36]. Poisson's ratio $v$ is greater than 0.26 for ductile solids and $v < 0.26$ for brittle materials. In view of Poisson's ratio, the Mo-based boride MoAlB is again identified as brittle material. Again, Poisson's ratio can predict the nature of chemical bonding in solids. Poisson's ratio $v$ is found to be around 0.10 for a covalent solid and around 0.33 for a metallic material, respectively [37]. The Poisson's ratio for MoAlB lies between these two characteristic values, indicating that the chemical bonding of this boride is a mixture of covalent and metallic in nature. A low Poisson's ratio is due to the directional bonding, which enhances the shear modulus and restricts the motion of dislocations, thereby increasing a material's hardness. A material is said to be hard if it is able to resist the plastic deformation. It involves preventing the nucleation and movement of dislocations, an irretrievable change in the structure, and is connected to both elastic and plastic processes in a solid. A material having short covalent bonding has a tendency of minimizing such dislocations and a material containing more delocalized bonds tolerates them [38]. Accordingly, diamond, the hardest material known to date, having covalent carbon–carbon bonding that shows high directionality with high strength. On the other hand, metallic compounds hold non-directional metallic bonds and are usually soft and ductile due to having a "sea of electrons".

*Elastic anisotropy*

Elastic anisotropy is an inherent characteristic of crystal solids. In addition to anisotropy of thermal expansion, elastic anisotropy involves introducing microcracks in crystals [39]. For this reason, the elastic anisotropy of crystals helps us to understand this property and uncovers the mechanisms, which can improve their durability. Naturally, all the known crystals exhibit elastic anisotropy and an accurate description of such anisotropic behavior are, therefore, important for its implication in crystal physics and engineering science. The anisotropy in orthorhombic crystals is due to shear anisotropy and anisotropy of linear bulk modulus. The shear anisotropic factors are very important, which can quantify the anisotropy in the bonding between atoms in different planes. The shear anisotropy factors for orthorhombic systems [40] are

$$A_1 = \frac{4C_{44}}{C_{11} + C_{33} - 2C_{13}}$$

for the (100) shear planes between the $\langle 011 \rangle$ and $\langle 010 \rangle$ directions,

$$A_2 = \frac{4C_{55}}{C_{22} + C_{33} - 2C_{23}}$$

for the (010) shear planes between the ⟨101⟩ and ⟨001⟩ directions, and

$$A_3 = \frac{4C_{66}}{C_{11} + C_{22} - 2C_{12}}$$

for the (001) shear planes between the ⟨011⟩ and ⟨010⟩ directions.

The deviation of the values of $A_1$, $A_2$, and $A_3$ from unity predicts the elastic anisotropy in crystals. In addition, any values of these factors smaller or greater than unity estimate the level of elastic anisotropy acquired by the crystals and indicate the in-plane and out-of-plane inter-atomic interactions differ from each other. The calculated values of these parameters are given in Table 3, indicating that the ternary boride MoAlB is elastically anisotropic and its in-plane and out-of-plane inter-atomic interactions differ significantly from each other. A more convenient measure of elastic anisotropy in polycrystalline solids is introduced with the percentage anisotropy in compressibility and shear [41]. These two factors are given successively as follows:

$$A_B = \frac{B_V - B_R}{B_V + B_R} \times 100\%$$

and

$$A_G = \frac{G_V - G_R}{G_V + G_R} \times 100\%$$

Here, $B$ and $G$ indicate the bulk and shear moduli and their subscripts $V$ and $R$ represent the Voigt and Reuss limits, respectively. When the values of both $A_B$ and $A_G$ are zero the materials are completely isotropic and the values greater than zero an increasing level of anisotropy is associated with the crystals. As well, a value of 100% signifies the highest possible anisotropy possessed by a crystal. It is evident from Table 3 that the anisotropy in compression is small but it is significant in shear.

Now, we want to concentrate on anisotropy of linear bulk modulus through considering the linear bulk modulus $B_a$, $B_b$, and $B_c$ along the a, b and c crystallographic axes, respectively. These parameters are defined as

$$B_a = a\frac{dP}{da} = \frac{\Lambda}{1 + \alpha + \beta}$$

$$B_b = b\frac{dP}{db} = \frac{B_a}{\alpha}$$

$$B_c = c\frac{dP}{dc} = \frac{B_a}{\beta}$$

where $\alpha$ and $\beta$ represent the relative change of the b and c axes as a function of the deformation of the a axis and defined for orthorhombic systems as follows:

$$\alpha = \frac{(C_{11} - C_{12})(C_{33} - C_{13}) - (C_{23} - C_{13})(C_{11} - C_{13})}{(C_{33} - C_{13})(C_{22} - C_{12}) - (C_{13} - C_{23})(C_{12} - C_{23})}$$

$$\beta = \frac{(C_{22} - C_{12})(C_{11} - C_{13}) - (C_{11} - C_{12})(C_{23} - C_{12})}{(C_{22} - C_{12})(C_{33} - C_{13}) - (C_{12} - C_{23})(C_{13} - C_{23})}$$

The parameter $\Lambda$ for orthorhombic crystals is

$$\Lambda = C_{11} + 2C_{12}\alpha + C_{22}\alpha^2 + 2C_{13}\beta + C_{33}\beta^2 + 2C_{23}\alpha\beta$$

Then, the anisotropies of the bulk modulus along the a and c axes with respect to the b axis can be expressed as

$$A_{B_a} = \frac{B_a}{B_b} = \alpha$$

and

$$A_{B_c} = \frac{B_c}{B_b} = \frac{\alpha}{\beta}$$

The calculated anisotropy of linear bulk moduli are also listed in Table 3. The unit value of these factors indicates isotropic elastic behavior of crystals and a value different from unity refers to a degree of elastic anisotropy. It is seen that the anisotropy of linear bulk modulus along $c$ axis is more significant than that along $a$ axis.

**Table 3**. The shear anisotropic factors $A_1$, $A_2$ and $A_3$, percentage anisotropy factors $A_B$ and $A_G$ and anisotropy of linear bulk moduli along a and c axes $A_{B_a}$ and $A_{B_c}$ for MoAlB.

| Compounds | $A_1$ | $A_2$ | $A_3$ | $A_B$ (%) | $A_G$ (%) | $A_{B_a}$ | $A_{B_c}$ | Ref. |
|---|---|---|---|---|---|---|---|---|
| MoAlB | 1.67 | 1.33 | 1.75 | 0.48 | 3.14 | 1.31 | 1.42 | This |
| ZrPtSi | 1.30 | 1.18 | 1.37 | 0.09 | 0.95 | -- | -- | [26] |
| TiPtSi | 1.25 | 1.10 | 1.22 | 0.52 | 0.66 | -- | -- | [26] |

*Debye temperature*

Debye temperature, as a fundamental parameter, leads to estimates many physical properties of solid materials namely, melting temperature, specific heat, lattice vibration, thermal conductivity, and thermal expansion. In addition, the Debye temperature has a relation with the vacancy formation energy in metals. To determine this parameter one standard method based on elastic properties is used frequently [42]. In this method the Debye temperature is calculated from the following equation using average sound velocity:

$$\theta_D = \frac{h}{k_B}\left[\left(\frac{3n}{4\pi}\right)\frac{N_A\rho}{M}\right]^{1/3} v_m$$

where $h$ refers Planck's constant, $k_B$ is symbolized for Boltzmann's constant, $N_A$ is Avogadro's number, $\rho$ denotes mass density, $M$ is the molecular weight and $n$ is the number of atoms in the molecule. The average sound velocity $v_m$ in the polycrystalline material is determined by

$$v_m = \left[\frac{1}{3}\left(\frac{1}{v_l^3} + \frac{2}{v_t^3}\right)\right]^{-1/3}$$

where $v_l$ and $v_t$ are the longitudinal and transverse sound velocities in the polycrystalline material and are obtained using the polycrystalline shear modulus $G$ and the bulk modulus $B$ as follows

$$v_l = \left[\frac{3B + 4G}{3\rho}\right]^{1/2}$$

and

$$v_t = \left[\frac{G}{\rho}\right]^{1/2}$$

The calculated Debye temperature $\theta_D$ along with sound velocities $v_l$, $v_t$, and $v_m$ are presented in Table 4. As a general rule, a higher Debye temperature is associated with a higher phonon thermal conductivity. The calculated Debye temperature as well as sound velocities for MoAlB is higher than those of two ternary silicides ZrPtSi and ZrPtSi. It implies that the Mo-based boride is thermally more conductive than these two silicide ternaries.

**Table 4**. Calculated density ($\rho$ in gm/cm$^3$), longitudinal, transverse and average sound velocities ($v_l$, $v_t$, and $v_m$ in km/s) and Debye temperature ($\theta_D$ in K) of MoAlB.

| Compounds | $\rho$ | $v_l$ | $v_t$ | $v_m$ | $\theta_D$ | Ref. |
|---|---|---|---|---|---|---|
| MoAlB | 6.33 | 7.95 | 4.77 | 5.28 | 693 | This |
| ZrPtSi | 10.76 | 5.62 | 3.35 | 3.71 | 437 | [26] |
| TiPtSi | 10.16 | 5.80 | 3.41 | 3.78 | 459 | [26] |

### *3.3 Electronic properties*

*Band structure and DOS*

The study of electronic band structure is useful to explain many physical properties such as optical spectra and transport properties of solids. The calculated energy band structure for MoAlB along the high-symmetry points in the *k*-space is shown in Fig. 2. The horizontal red line is drawn as the Fermi level. A few valence bands cross this line and overlap with conduction bands. For this reason this new ternary boride should exhibit metallic conductivity.

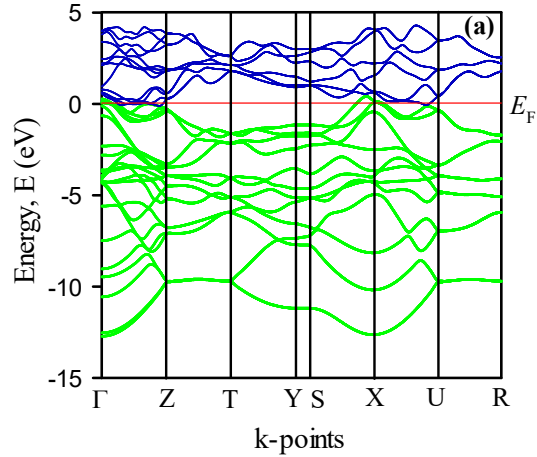

**Fig. 2.** Electronic band structure along with high symmetry directions of MoAlB.

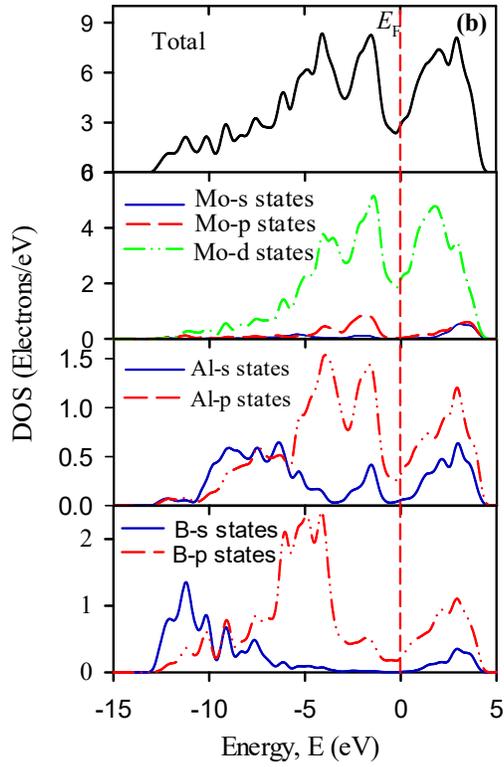

**Fig. 3.** Total and partial electronic density of states (DOS) of MoAlB.

The total and partial density of states (DOS) for MoAlB is presented in Fig. 3, in which the vertical dotted line represents the Fermi level, $E_F$ set to 0 eV. The DOS at $E_F$ has a value of 2.9 states/unit cell-eV. A characteristic feature of DOS profile of MoAlB is that there is a pseudo-gap near the Fermi level that serves as the borderline between the bonding and anti-bonding states and is cooperative for stabilizing the structure of MoAlB. In the plot, it is seen that wide valence bands are originated and spreading from Fermi level to ~13 eV. The valence bands contain many peaks in its structure. The lowest peaks are mainly due to s-states of B that indicates the formation of strong covalent B-B bonding in MoAlB. The peaks in the middle part of the valence bands originate from the p-states of B and Al as well as d-states of Mo. Al s-state also contributes a little bit to form these valence bands. These hybridizations lead to form the covalent bonding of B with Al and Mo as well as a bonding between Mo and Al. The highest valence bands with highest peak occur due to hybridization of B p, Al p and Mo d electrons. The second highest peak near the Fermi level arises from the s- and p-states of Al and p- and d-states of Mo. At the Fermi level, the DOS originates mainly from the d-electrons of Mo. This d-resonance in the Fermi level as well as the finite value of the DOS at $E_F$ indicates the metallic conductivity of the ternary compound MoAlB. The energy region above the Fermi level is dominated by Mo d and Al p states.

*Mulliken atomic and bond overlap populations*

The allocation of electrons in several fractional ways among the different parts of chemical bonds can be explained with the Mulliken population analysis. Segall *et al.*[43] also observed a good relation of the overlap population with covalency of bonding and bond strength. Ching and Xu [44] stated that the overlap population may be a convenient method to quantify the strength of bonding in first-principles calculations. But, a problem arises due to the use of a plane-wave basis set in the first-principles formalism within the CASTEP code. Because, the extended basis states never provide a natural way to quantify the local atomic properties. To solve this issue Sanchez-Portal *et al.* [45, 46] described a method in which a projection of plane-wave states onto a linear combination of atomic orbitals (LCAO) is used. This technique is frequently used to calculate the atomic charges and the bond populations via Mulliken population analysis [43, 47]. In this analysis, the Mulliken charge coupled with a particular atom $\alpha$ can be expressed as:

$$Q(\alpha) = \sum_k w_k \sum_\mu^{on\ \alpha} \sum_\nu P_{\mu\nu}(k) S_{\mu\nu}(k)$$

and the overlap population between two atoms $\alpha$ and $\beta$ is

$$P(\alpha\beta) = \sum_k w_k \sum_\mu^{on\ \alpha} \sum_\nu^{on\ \beta} 2P_{\mu\nu}(k) S_{\mu\nu}(k)$$

where $P_{\mu\nu}$ is elements of the density matrix and $S_{\mu\nu}$ is the overlap matrix. The effective valence charge and Mulliken atomic population are two important factors because of their role for realizing the nature of chemical bonding in crystalline solids. The difference between the formal ionic charge and Mulliken charge on the anion species in a crystal is defined as the effective valence charge. It determines the strength of a bond either as covalent or ionic. An ideal ionic bond occurs when the effective valence has exactly zero value. An atom associated with a non-zero effective valence charge prefers covalent bonding and the level of covalency will be estimated by the degree of deviation from zero. Table 5 lists the calculated effective valence that indicates the existence of prominent covalency in chemical bonding inside the ternary compound MoAlB.

**Table 5**. Mulliken atomic and bond overlap population of MoAlB

| Mulliken atomic population | | | | | | | Mulliken bond overlap population | | | |
|---|---|---|---|---|---|---|---|---|---|---|
| Atoms | s | p | d | Total | Charge (e) | Effective valence charge (e) | Bond | Bond number $n^\mu$ | Bond length $d^\mu$ (Å) | Bond overlap population $P^\mu$ |
| B | 0.95 | 2.53 | 0.00 | 3.49 | - 0.49 | -- | B-B | 2 | 1.80931 | 1.38 |
| | | | | | | | B-Al | 4 | 2.32046 | 0.15 |
| Al | 0.85 | 1.87 | 0.00 | 2.72 | 0.28 | 2.72 | B-Mo | 4 | 2.36965 | 0.66 |
| | | | | | | | Al-Al | 2 | 2.66718 | 0.95 |
| Mo | 2.11 | 6.47 | 5.21 | 13.79 | 0.21 | 5.79 | Al-Mo | 4 | 2.71119 | 0.73 |

The bond overlap population is calculated for MoAlB with distance cutoff for bond population 3 Å and is presented in Table 5. The bond overlap population provides quantitative information, which can be used as a measure of bonding and antibonding states. Positive and negative population indicates bonding and antibonding states, respectively. No significant interaction is observed between the electronic populations of two atoms when the overlap population is associated with a value close to zero. On the other hand, a high level of covalency is observed in a chemical bond that contains a high overlap population. The order of covalency level found in chemical bonds can be expressed as B-B > Al-Al > Al-Mo > B-Mo > B-Al. The reverse of this order will indicate the order of ionicity level of the chemical bonds. Hence, B-Al bond is more ionic than other bonds.

*3.4 Vickers hardness and bonding nature*

Material's hardness is its ability to resist plastic deformation. F. M. Gao [48] formulated an equation for estimating the Vickers hardness of non-metallic compounds by means of Mulliken population within first-principles method. This formula is not suitable for crystals having partial metallic bonding. Because metallic bonding is delocalized and there exists no crucial relation between metallic bonding and hardness [49]. Gou et al. [50] reformulated their equation considering a correction due to delocalized metallic bonding for compound having partial metallic bond and expressed it as follows:

$$H_v^\mu = 740(P^\mu - P^{\mu'})(v_b^\mu)^{-5/3}$$

where $P^\mu$ is the Mulliken overlap population of the $\mu$-type bond, $P^{\mu'}$ is the metallic population and is evaluated from the unit cell volume $V$ and the number of free electrons in a cell $n_{free} = \int_{E_P}^{E_F} N(E)dE$ as $P^{\mu'} = n_{free}/V$, and $v_b^\mu$ is the volume of a bond of $\mu$-type, which is calculated via the bond length $d^\mu$ of type $\mu$ and the number of bonds $N_b^\nu$ of type ν per unit volume as $v_b^\mu = (d^\mu)^3/\sum_\nu[(d^\mu)^3 N_b^\nu]$. The geometric average of all bonds' hardness gives the total hardness of a complex multiband crystal and can be expressed as [51, 52]:

$$H_V = [\prod^\mu (H_v^\mu)^{n^\mu}]^{1/\sum n^\mu}$$

where $n^\mu$ refers the number of bond of type $\mu$ composing the real multiband crystals. The calculated Vickers hardness for world's first corrosion-resistant boride MoAlB is listed in Table 6 along with relevant quantities. The calculated value of Vickers hardness is 11.6 GPa, which lies between the ranges of measured values (10.1 – 13.6 GPa). Kota *et al.* [4] measured with different loads on cross-section and found hardness around 10.6 ± 0.3 GPa that is very close to the value (10.3 ± 0.2 GPa) measured on B-planes of MoAlB single crystals [2]. Ade *et al.* [1] estimated the hardness on different crystal planes and obtained the values ranging from 11.4 – 13.6 GPa. It is obvious that Mo-based boride MoAlB is softer in comparison to many other harder borides namely, MoB ($H_V$ = 23 GPa) [53] and MoB$_2$ ($H_V$ = 21 – 27 GPa) [54, 55].

**Table 6** Calculated bond and Vickers hardness $H_v^\mu$, $H_v$ (in GPa) of MoAlB along with bond number $n^\mu$, bond length $d^\mu$ (Å), bond volume $v_b^\mu$ (Å$^3$) and bond as well as metallic populations $P^\mu$, $P^{\mu'}$.

| Bond | $n^\mu$ | $d^\mu$ | $P^\mu$ | $P^{\mu'}$ | $v_b^\mu$ | $H_v^\mu$ | $H_v$ |
|---|---|---|---|---|---|---|---|
| B–B | 2 | 1.80931 | 1.38 | 0.0048 | 3.5704 | 122.012 | 11.6 |
| B–Al | 4 | 2.32046 | 0.15 | 0.0048 | 7.5318 | 3.713 | |
| B–Mo | 4 | 2.36965 | 0.66 | 0.0048 | 8.0210 | 15.085 | |
| Al–Al | 2 | 2.66718 | 0.95 | 0.0048 | 11.4376 | 12.047 | |
| Al–Mo | 4 | 2.71119 | 0.73 | 0.0048 | 12.0132 | 8.516 | |

The metallicity of bonding can be evaluated from $f_m = P^{\mu'}/P^\mu$ [56, 57]. The calculated values for the bonds B–B, B–Al, B–Mo, Al–Al and Al–Mo are found to be 0.00348, 0.03200, 0.00727, 0.00505 and 0.00657, respectively. The ranking of metallicity of the bonds are B–Al > B–Mo > Al–Mo > Al–Al > B–B. The lowest value of bond population corresponds to the highest value of metallicity.

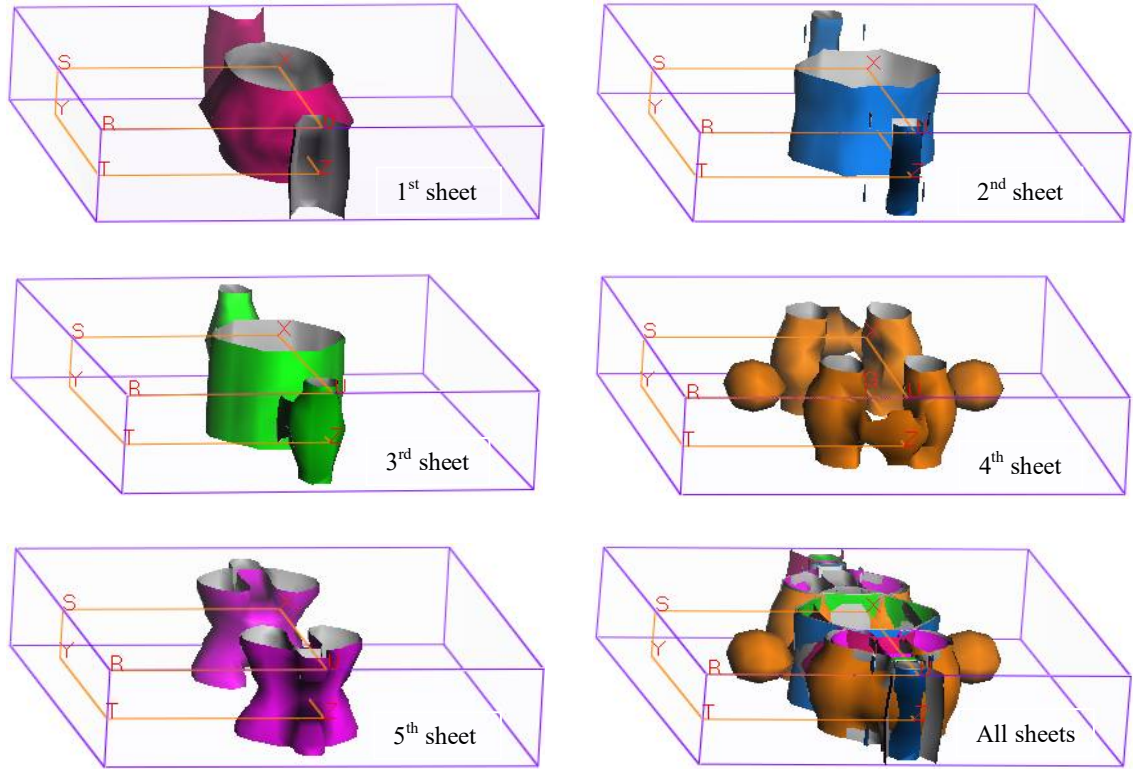

**Fig. 4.** Fermi surface topology of MoAlB with individual sheets.

## *3.5 Fermi surface and charge density*

The calculated Fermi surface is shown in Fig. 4 along with the individual sheets. Both electron and hole like sheets are seen in Fermi surface topology. First three sheets are centered along Γ–X direction. The first sheet looks like the container of a sand-cement mixing machine. It contains two additional parts parallel to Z–U directions and like a panel of a corrugated metal sheet. One part situates on an opposite position of another. The second sheet is cylindrical with hexagonal cross-section. It has also two parts parallel to Z–U directions and like tubes with square cross-section. One tube takes opposite position of another. The third sheet is almost similar to the second sheet. The main difference is the shape of the tubes. These tubes have two narrow ends and swollen belly. The fourth sheet is very complicated and it consists of four tubes with narrow ends and swollen belly. These four tubes form a rectangular shape and each tube is centered along one edge of the rectangular. The same-sided tubes are connected to each other. Each of two tubes situated at the ends of a diagonal of the bottom face of the rectangular has a spheroid like appendix to outer side. The fifth and final sheet has two sand-watch shaped tubes connected along X–U direction. The both ends of the tube have nose shaped appendices being one inside and another outside of the tubes. The whole sheets form together a complex topology of the Fermi surface. The low-dispersive Mo 4d-like bands play role to form the Fermi surface of MoAlB. It is also responsible for the electrical conductivity of the compound.

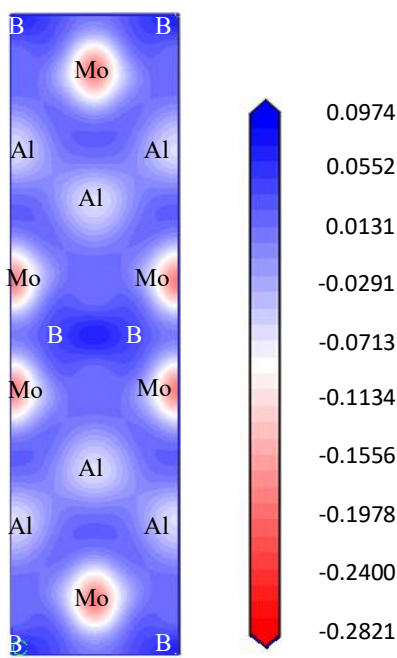

**Fig. 5.** Charge density distribution map of ternary boride MoAlB.

The electron charge density map reveals the electron densities associated with the chemical bonds. It includes areas of positive as well as negative charge densities and indicates both accumulation and depletion of electronic charges. From the charge density map, the covalent bonds can be identified via the accumulation of charges between two atoms. The existence of ionic bonds can be predicted from a negative and positive charge balance at the atom positions [58]. The valence electronic charge density map (in the units of e/Å$^3$) for MoAlB is shown in Fig. 4 along (110) crystallographic plane. The scale adjacent to the map estimates the intensity of electronic charge density: The red and blue colors in the scale indicate the low and high density of electronic charge, respectively. From the Fig.5, it is evident that strong covalent bonding B-B exists between B-atoms. There exists another covalent bonding Al-Al between Al-atoms, which leads to form the al-layer in the compound. The strength of this bonding is slightly less than that of B-B bonding. The accumulation of charge also indicates the covalent bonding between B-Mo, Al-Mo, and B-Al.

4. Conclusions

In this paper, the structural, elastic and electronic properties of MoAlB have been investigated by means of first-principles DFT calculations. The calculated lattice parameters are in good agreement with the experimental values. The ternary boride MoAlB is more compressible along the *b*-axis than that of along the other two axes. The analysis of bulk modulus to shear modulus ratio, Poison's ratio and Cauchy pressure predict that MoAlB should be brittle in nature. The calculated shear anisotropy factors assure elastic anisotropy in MoAlB and its in-plane and out-off-plane inter-atomic interactions differ notably from each other. The studied boride exhibits small anisotropy in compression but large in shear. Moreover, the anisotropy of linear bulk modulus along *c* axis is seen to be more significant than that along *a*-axis. The comparatively high Debye temperature of MoAlB indicates its high thermal conductivity. The calculated band structure and DOS reveal the metallic conductivity of MoAlB. The investigated DOS, Mulliken populations and electron charge density distribution suggest that the chemical bonding in MoAlB is essentially covalent in nature. In comparison to many other hard borides, MoAlB is soft. The theoretically constructed Fermi surface topology is quite complicated and it contains both electron and hole like sheets. The low-dispersive Mo 4d-like bands play prominent role to the formation of the Fermi surface of MoAlB.